\documentclass[journal=jpccck,manuscript=article]{achemso}

\usepackage{chemformula} 
\usepackage[T1]{fontenc} 
\usepackage{amssymb,amsmath,graphics,graphicx,float,braket}
\usepackage{epstopdf}
\usepackage{array,multirow,setspace}
\usepackage{subfig}
\usepackage[autostyle]{csquotes}

\title{Probing Photoexcited Charge Carrier Trapping and Defect Formation in Synergistic Doping of \ch{SrTiO3}}
\author{Namitha Anna Koshi}
\affiliation{Indo-Korea Science and Technology Center (IKST), Yelahanka, Bengaluru, India}
\author{Dharmapura H K Murthy}
\affiliation{Materials Science and Catalysis Division, Poornaprajna Institute of Scientific Research, Devanahalli, Bengaluru, India}
\author{Sudip Chakraborty}
\affiliation{Indian Institute of Technology (IIT) Indore, India}
\author{Seung-Cheol Lee}
\affiliation{Indo-Korea Science and Technology Center (IKST), Yelahanka, Bengaluru, India}
\email{seungcheol.lee@ikst.res.in}
\author{Satadeep Bhattacharjee}
\affiliation{Indo-Korea Science and Technology Center (IKST), Yelahanka, Bengaluru, India}
\email{s.bhattacharjee@ikst.res.in}

\abbreviations{IR,NMR,UV}

\begin{document}

\begin{tocentry}
\includegraphics[scale=0.40]{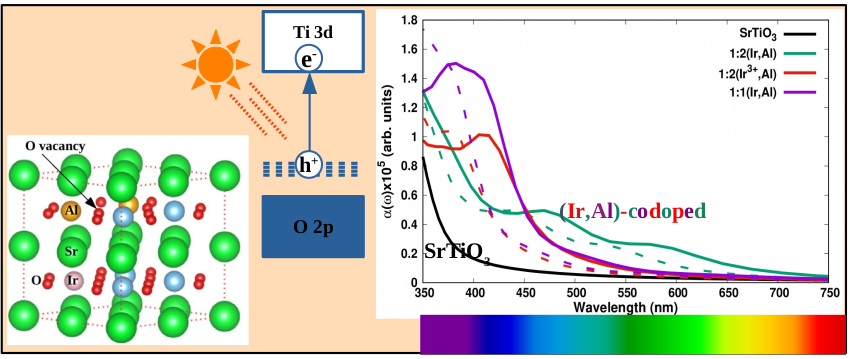}
\end{tocentry}

\begin{abstract}
Strontium titanate (\ch{SrTiO3}) is widely used as a promising photocatalyst due to its unique band edge alignment with respect to the oxidation and reduction potential corresponding to oxygen evolution reaction (OER) and hydrogen evolution reaction (HER). However, further enhancement of the photocatalytic activity in this material could be envisaged through the effective control of oxygen vacancy states. This could substantially tune the photoexcited charge carrier trapping under the influence of elemental functionalization in \ch{SrTiO3}, corresponding to the defect formation energy. The charge trapping states in \ch{SrTiO3} decrease through the substitutional doping in Ti sites with p-block elements like Aluminium (Al) with respect to the relative oxygen vacancies. With the help of electronic structure calculations based on density functional theory (DFT) formalism, we have explored the synergistic effect of doping with both Al and Iridium (Ir) in \ch{SrTiO3} from the perspective of defect formation energy, band edge alignment and the corresponding charge carrier recombination probability to probe the photoexcited charge carrier trapping that primarily governs the photocatalytic water splitting process. We have also systematically investigated the ratio-effect of Ir:Al functionalization on the position of acceptor levels lying between Fermi and conduction band in oxygen deficient \ch{SrTiO3}, which governs the charge carrier recombination and therefore the corresponding photocatalytic efficiency. 
\end{abstract}

\section{1. Introduction}
\label{section 1}
Photocatalytic water splitting is a process by which water molecules are split into \ch{H2} and \ch{O2} using solar energy on the surface of a catalyst. \ch{H2}, one of the products, is a clean and renewable source of energy. The oxide perovskites are a type of inorganic semiconductor photocatalysts, which are known to show extremely high stability than N-containing or S-containing photocatalysts. Also, long term stability in water is one of the many factors that influence photocatalytic efficiency and titanates are known to be stable in aquatic environments \cite{Fujishima1972, Ni2007, Kawasaki2012}. In general, oxide perovskites are restrained by their wide band gaps. So they have to be engineered to use visible light, which constitutes 43$\%$  of the incoming solar energy \cite{Zou2001}. For efficient utilization of visible region of the solar spectrum, a photocatalyst should have band gap, $E_{g}$ between 1.8 and 2.2 eV \cite{Nowotny2005, Razavi2017}, though 1.23 eV is the theoretical lower limit for water photolysis. Also, the valence band maximum (VBM) and the conduction band minimum (CBM) of the photocatalyst should straddle the redox potentials of water.\\

To determine the potential utility of a semiconductor for photocatalysis, it is essential to understand its electronic structure. For this purpose, most of the earlier theoretical investigations on wide band gap semiconductors used local density approximation (LDA) and generalized gradient approximation (GGA). The inadequacy of LDA and GGA to determine band gaps of semiconductors and insulators have already been discussed in the past and is mainly ascribed to the missing discontinuity in local exchange-correlation potential and self-interaction error \cite{Perdew1983, Sham1983, Stadele1997}. In addition to the band gap underestimation, LDA and GGA, also underestimates the binding energy of semi-core d states which in turn, leads to overestimation of their hybridization with the anion-p valence states, which amplifies the effects of p-d coupling \cite{Janotti2006, Ma2011, Ma2013}. This large p-d coupling shifts the valence band maximum upward which results in further reduction of band gap. Sometimes, it leads to the closing of band gap as in CdO and InN, thereby predicting wrong metallic ground state \cite{Janotti2006}. For a transition metal oxide like ZnO, advanced hybrid functional (HSE) and GW implementations, did not fully improve the underestimation of band gap and under-binding of Zn 3d states \cite{Uddin2006, Lany2010}. But in many reports, hybrid functionals have been used, which are computationally very demanding, for the study of \ch{TiO2} and \ch{SrTiO3}. DFT+U method is an effectively better choice to treat the band gap and binding energy underestimation of semi-core d states in these semiconductors, as it requires less computational effort \cite{Ma2011, Uddin2006, Sheetz2009}. This is done by adding an orbital dependent term to the potential. We select the appropriate value of U by matching the electronic structure obtained with DFT+U to that of Heyd-Scuseria-Ernzerhof hybrid functional (HSE06). A similar approach has been followed in determining the electronic structure of molecular magnet, Co-phthalocyanine \cite{bhattacharjee2010}.\\

\ch{SrTiO3} is an ultraviolet (UV) active photocatalyst with a wide band gap of 3.2 eV \cite{Cardona1965, Wagner1980}. Though the band edges align with water redox potentials, the band gap must be tuned to make it a potential visible light active photocatalyst. Introducing dopants has been proposed as an effective method to reduce the band gap and transition metal (TM) doping is known to extend the spectral response to visible region \cite{Niishiro2010, Asai2014, Kudo2019}. Generally, when \ch{SrTiO3} is doped with a TM at cationic site, it introduces discrete localized states in the forbidden region which may act as recombination centers for charge carriers, which leads to reduction in the number of photo-excited carriers \cite{Kumar2020, Pelaez2012}. The position of the dopant states between the valence band and conduction band plays an important role in determining photocatalytic activity of a doped system \cite{Chen2012}. In some cases, codoping resulted in passivating the mid-gap localized states formed by TM dopants. Cationic (Cr, Ta)-codoping resulted in better activity than Cr-doped \ch{SrTiO3} by suppressing the formation of Cr$^{6+}$ ions which work as recombination centers \cite{Ishii2004}. Codopants can be at cation-anion, anion-anion or cation-cation sites. \\

We are investigating the effect of cation-cation codoping on \ch{SrTiO3} photocatalytic behavior. The dopants considered are Al and Ir. The effect of these monodopants on \ch{SrTiO3} have been studied theoretically and they are also experimentally synthesized \cite{Chen2012, Zhao2019, Kawasaki2014, Suzuki2018, Takata2020}. The n-type semiconductivity of \ch{SrTiO3} is due to the formation of Ti$^{3+}$ defect levels below the bottom of conduction band, a consequence of its intrinsically slightly oxygen deficient, non-stoichiometric character \cite{Takata2009}. 
Based on the results of Zhao \textit{et al.} \cite{Zhao2019}, Al-doping reduces the n-typeness of \ch{SrTiO3} and electron-hole recombination losses, thereby improving their photocatalytic properties. Also, using modified Al-doped \ch{SrTiO3} for water splitting, external quantum efficiency of upto 96\% was obtained at wavelengths between 350 and 360 nm \cite{Takata2020}. It is necessary to widen the quantum efficiency of photocatalysts over a large range of wavelengths to achieve better solar-to-hydrogen conversion efficiency. From previous reports on TM doping, it is understood that, though Rh and Ir belong to the same group in the periodic table, their photocatalytic responses are different. The photocatalytic activity of Rh-doped \ch{SrTiO3} is better than Ir-doped \ch{SrTiO3} but Ir-doped \ch{SrTiO3} is known to show absorption over a wide range of visible light \cite{Kawasaki2014}. Therefore, by codoping \ch{SrTiO3} with Ir and Al, we intend to reduce the n-type character of Ir-doped \ch{SrTiO3} and charge recombination losses and also, extend the spectral response of Al-doped \ch{SrTiO3} more into the visible region of sunlight. In this report, we have constructed a 2$\times$2$\times$2 supercell consisting of 40 atoms and introduced an oxygen vacancy by removing one of the oxygen atoms and further substituted dopants. To the best of our knowledge, there are few reports on Al-doped and Ir-doped \ch{SrTiO3} but not on (Ir,Al)-codoped \ch{SrTiO3}. Hence, we present the electronic and optical properties of (Ir,Al)-codoped \ch{SrTiO3} obtained using DFT+U formalism with the electronic structure of pristine, Al-doped, 1:2 (Ir,Al)-codoped \ch{SrTiO3} matched with that obtained employing hybrid functional. \\

\section{2. Computational details}
\label{section 2}
The electronic structure calculations are performed using projector augmented plane wave method implemented in Vienna ab-initio Simulation Package (VASP) \cite{Kresse1996, Kresse1999, Blochl1994}. The projector augmented wave (PAW) potentials represent the interactions between valence electrons and ionic core for all the species \cite{Blochl1994} and the exchange-correlation is treated by employing generalized gradient approximation (GGA) parameterized by Perdew-Burke-Ernzerhof (PBE) formalism \cite{Perdew1996}. In order to account for the inability of PBE-GGA to accurately calculate the band gap of semiconductors and insulators, we used the DFT+U approach as proposed by Dudarev \textit{et al.}\cite{Dudarev1998}. This is carried out by mapping the electronic structure of \ch{SrTiO3} to the known experimental reports. For k-point integration within the first Brillouin zone, a Monkhorst-Pack \cite{Monkhorst1976} mesh of 7$\times$7$\times$7 for primitive cell and 4$\times$4$\times$4 for supercells are considered. The energy convergence criterion is set to 10$^{-6}$ eV. All structures are relaxed until the residual forces are less than 0.04 eV/\AA{}. A cut-off energy of 500 eV is used for all calculations. The defect formation energy is calculated for monodoped and codoped systems. Although DFT+U do not give the correct energetics for metal oxides \cite{Basera2019}, we focus on the electronic structure which are described well by it. Also, spin-polarized calculations are carried out as some of the doped systems have unpaired electrons. The VASP data obtained after calculations are post-processed using VASPKIT \cite{Wang2019} and crystal structures are visualized using VESTA \cite{Momma2011}. Some of the electronic structure calculations are carried out using Heyd-Scuseria-Ernzerhof (HSE) hybrid functional \cite{Heyd2003,Krukau2006}. Here, 28\% Hartree-Fock mixing for exchange and standard value of 0.2 \AA{}$^{-1}$ for screening parameter gives band gap of pristine \ch{SrTiO3} close to that of experimental value.\\

\section{3. Results and Discussion}
\label{section 3}
\subsection{3.1 Geometry and Electronic structure}
\label{subsection 3.1}
\subsubsection{3.1.1 \ch{SrTiO3}} 
\label{subsubsection 3.1.1}
\ch{SrTiO3} has a cubic perovskite structure with space group Pm-3m. The relaxed bond length of Ti-O is 1.97 \AA{} and distance between Sr and nearest O is 2.79 \AA{}. The optimized lattice parameter of \ch{SrTiO3} is 3.95 \AA{} (using GGA). The bond length and lattice parameter values of \ch{SrTiO3} agrees with other ab-initio results using PBE-GGA potential \cite{Chen2012}. The electronic structure of \ch{SrTiO3} is studied in detail elsewhere \cite{Hamid2009, Chen2012}. We have obtained a band gap of 1.67 eV using PBE-GGA, which is far less than the experimental value of band gap (3.2 eV). This underestimation is due to the lack of sufficient self-interaction error correction present in the local exchange-correlation functional. To get a better band gap value and for a correct description of p-d hybridization, we use the DFT+U method. To investigate the effect of U on Ti-d and O-p orbitals in \ch{SrTiO3}, we consider a series of U values ranging from 4 to 10 eV. The details are given in supporting information.\\

For U$_{d}$(Ti) = 8 and U$_{p}$(O) = 8, \ch{SrTiO3} has a band gap of 3.13 eV, which is very close to the experimental value and the corresponding lattice parameter obtained is 4.00 \AA{}. The Ti-O bond length increases to 2.00 \AA{} with Hubbard U correction and the corresponding Sr-O length is 2.83 \AA{} which is comparable to GGA structure. Ma \textit{et al.} suggested U = 7 for description of O-2p states of oxide materials based on their work on Ag$_{3}$PO$_{4}$ \cite{Ma2011} and from other reports on SiO$_{2}$ \cite{Nolan2006}, MgO \cite{Nolan2005} and Auger spectroscopic studies of different oxides \cite{Ghijsen1988, Knotek1978}. This is in near agreement with our chosen value of U$_{p}$(O). The band structure and total density of states (TDOS) of \ch{SrTiO3} plotted using PBE-GGA and DFT+U are provided in supporting information (Figure SI2 and SI3). The results are consistent with reported literature \cite{Takizawa2009}. The projected density of states (PDOS) of the pristine \ch{SrTiO3} (2$\times$2$\times$2 cell - for the sake of comparison with doped systems) is shown in Figure 1. As can be seen from the PDOS, the VB is dominated by the O-p states and the CB is dominated by Ti-d states. Sr do not contribute much to the states near the Fermi level. The density of states is symmetric with respect to spin channels, which implies that pure \ch{SrTiO3} has no net magnetic moment. \\

Oxygen vacancy is one of the ubiquitous and important defects in oxide-based perovskites. Therefore, it is necessary to understand its effect on the electronic structure of \ch{SrTiO3}. To the 2$\times$2$\times$2 supercell, we introduce an oxygen vacancy by removing one of the oxygen atoms (1 in 24). The relaxed oxygen deficient \ch{SrTiO3} (\ch{Sr8Ti8O23}) is given in Figure 1(b).  Earlier reports suggest that the removal of one oxygen atom should introduce n-type carriers in \ch{SrTiO3} and the Fermi level shifts into the conduction band, giving rise to a semiconductor-to-metal transition \cite{Hamid2009, Wen2011}. In the present study, oxygen vacancy shifts the Fermi level into the CB  with the occupied Ti-O hybrid states lying below and at $E_{F}$, separated from the rest of the CB by a 0.5 eV gap. Both the references did not report this gap in CB. This could be because of the different exchange-correlation potential and first-principles code used. Since we are introducing U for O-p states, our results present the p-d hybridization between O and Ti better, which is overestimated in GGA \cite{Ma2011}. The loss of an oxygen ion must be accompanied by the creation of two Ti$^{3+}$ ions to keep charge neutrality \cite{Zhao2019}, therefore the states, we observe just below $E_{F}$ is due to the formation of Ti$^{3+}$. From this point onwards, we consider only oxygen deficient \ch{SrTiO3}.
\begin{figure}
\begin{center}
\subfloat[]{\includegraphics[scale=0.38]{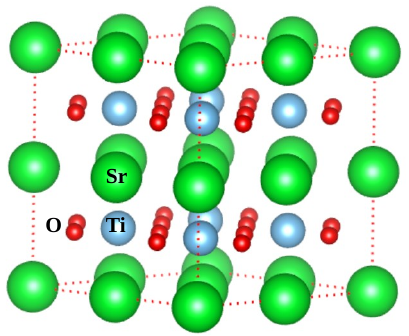}}
\subfloat[]{\hspace{1.5cm}\includegraphics[scale=0.4]{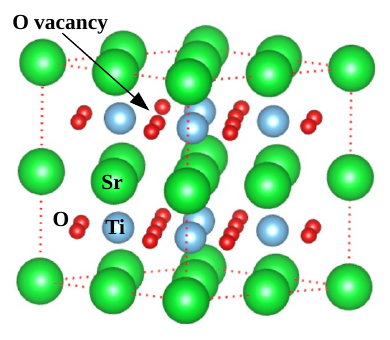}}
\subfloat[]{\hspace{1.5cm}\includegraphics[scale=0.35]{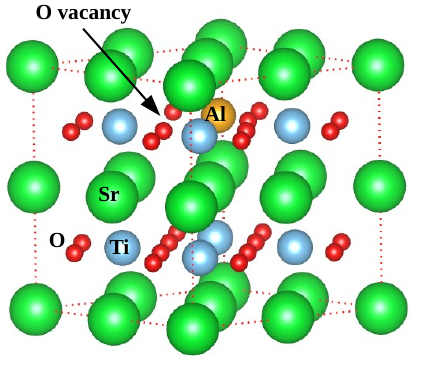}}\\
\subfloat[]{\includegraphics[scale=0.4]{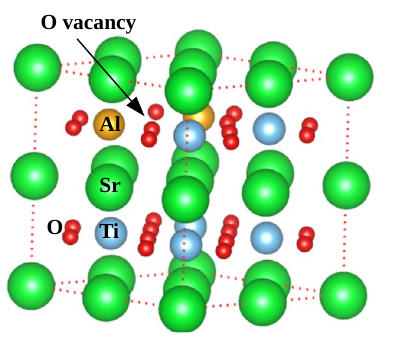}}
\subfloat[]{\hspace{1.5cm}\includegraphics[scale=0.4]{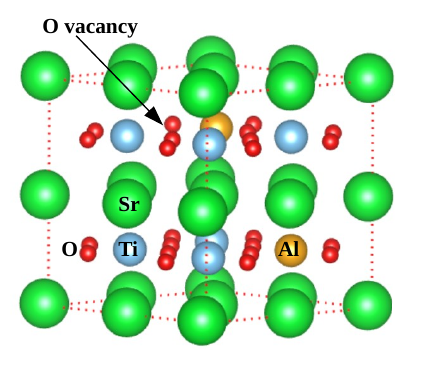}}
\subfloat[]{\hspace{1.5cm}\includegraphics[scale=0.4]{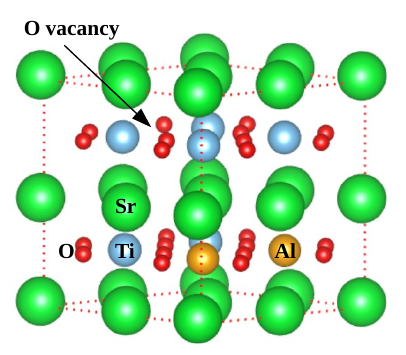}}\\
\subfloat[]{\hspace{-1cm}\includegraphics[scale=0.45]{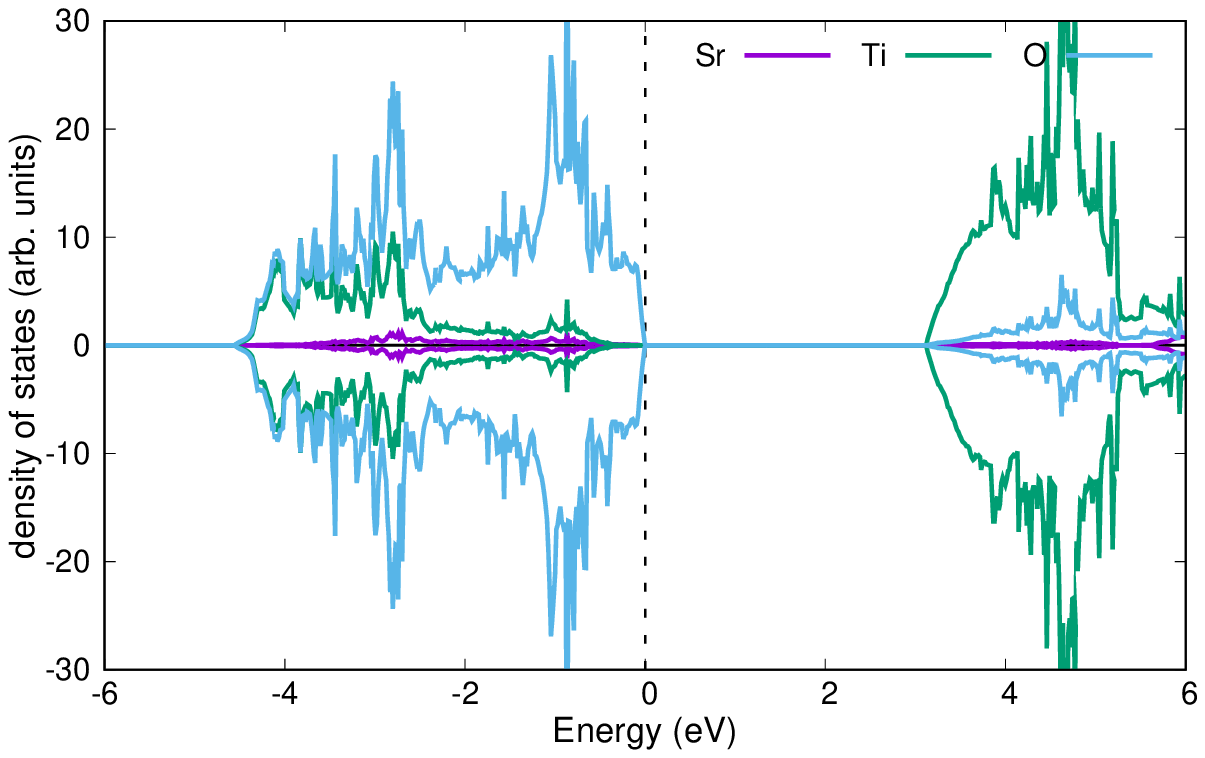}} 
\subfloat[]{\includegraphics[scale=0.45]{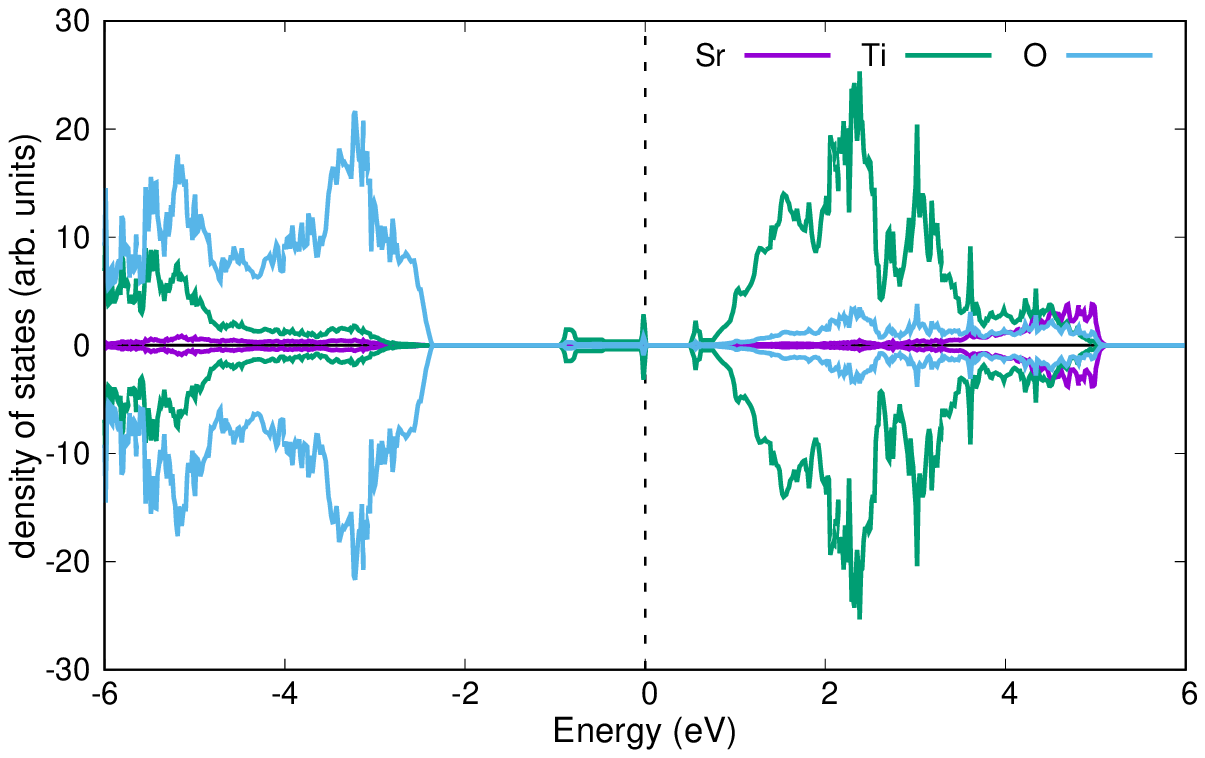}} 
\subfloat[]{\includegraphics[scale=0.45]{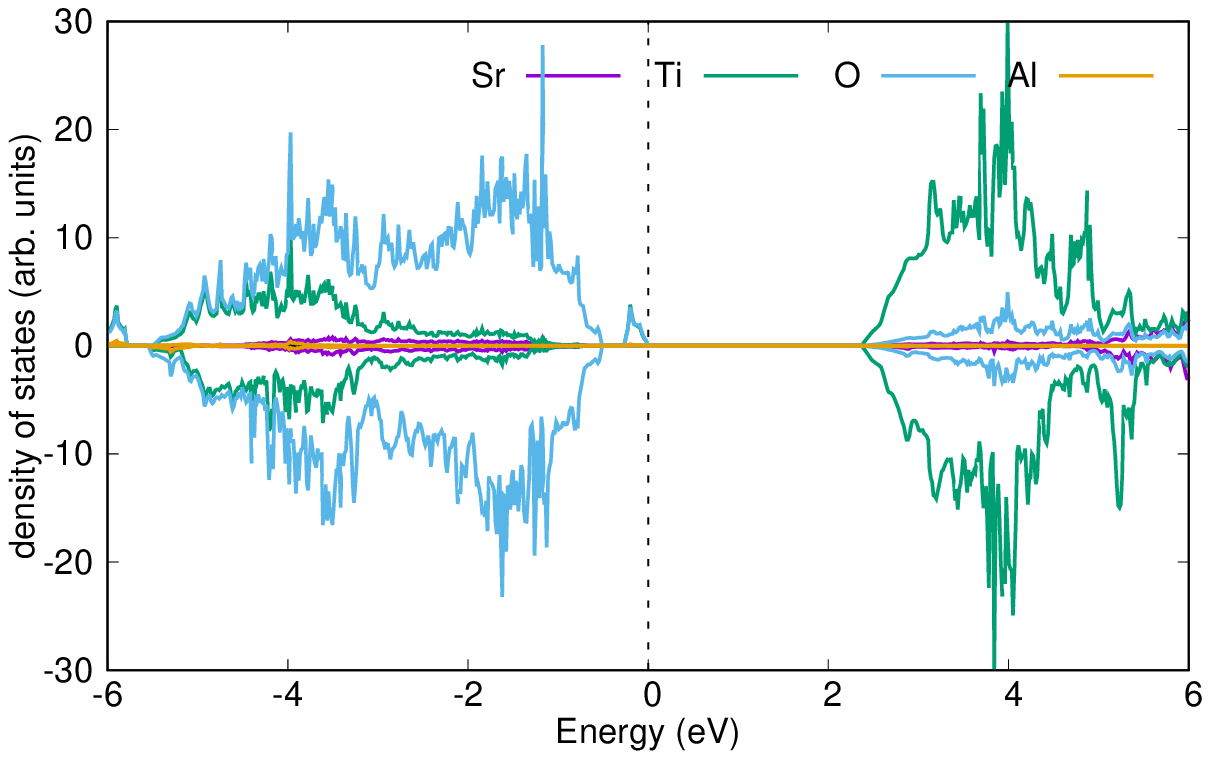}} \\
\subfloat[]{\hspace{-1cm}\includegraphics[scale=0.45]{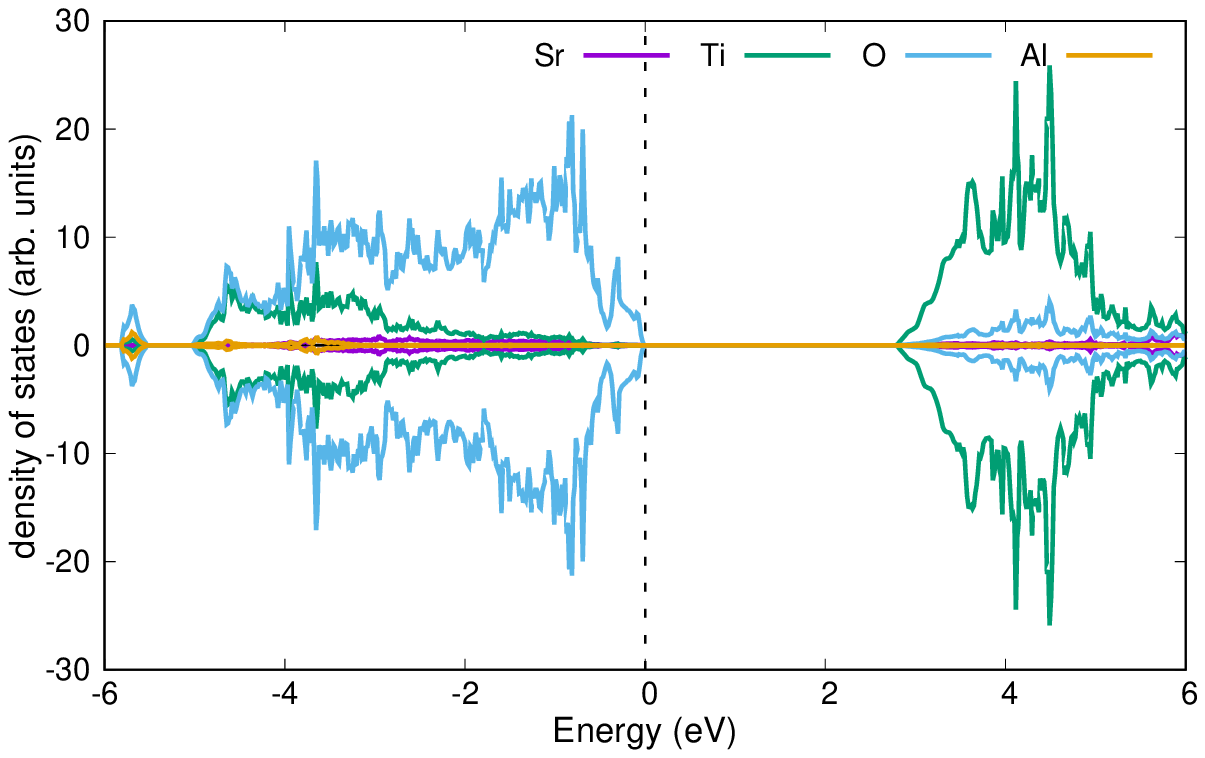}} 
\subfloat[]{\includegraphics[scale=0.45]{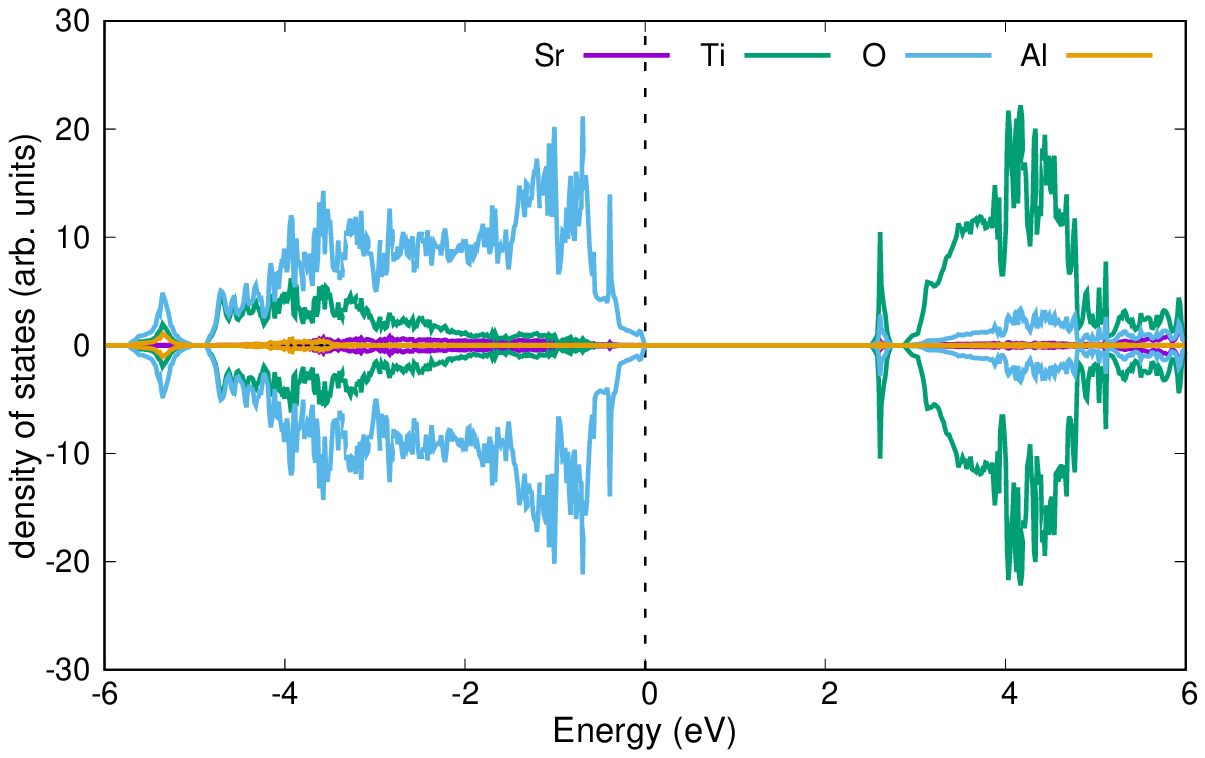}} 
\subfloat[]{\includegraphics[scale=0.45]{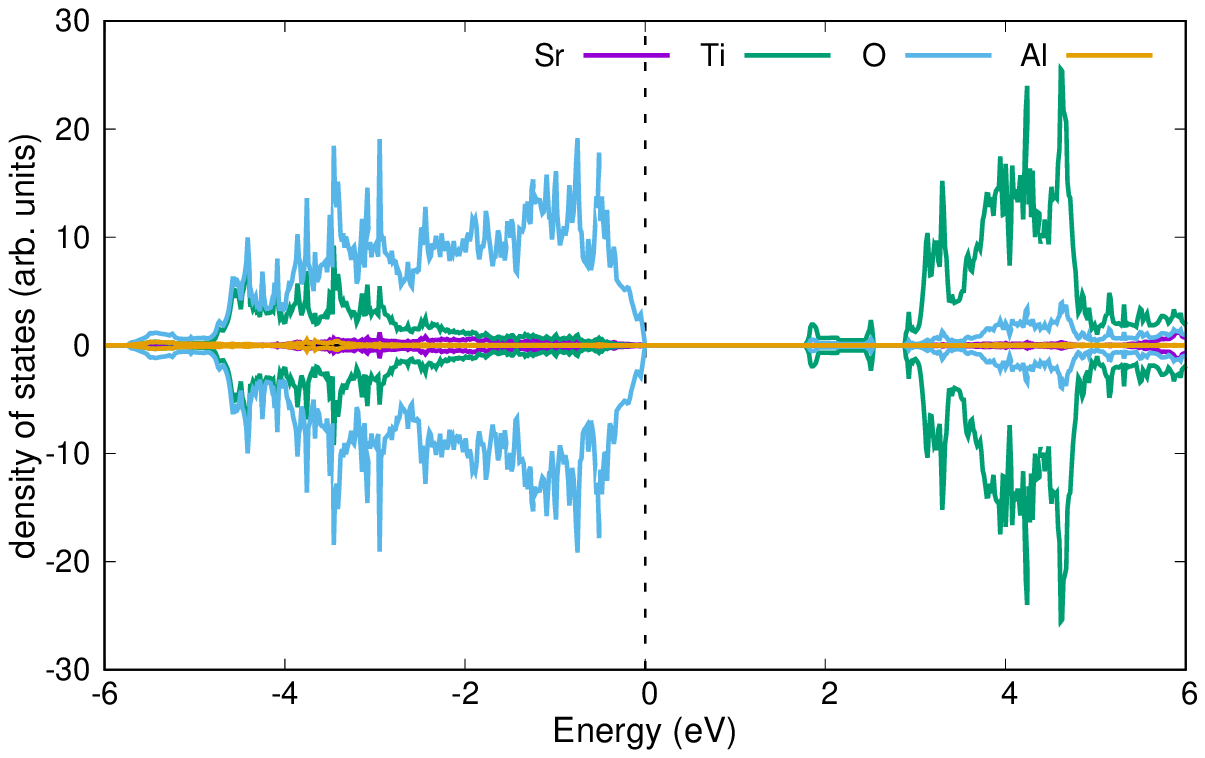}}
 \end{center}
 \caption{Relaxed structure of (a) 2$\times$2$\times$2 super cell - \ch{Sr8Ti8O24}, (b) oxygen deficient \ch{SrTiO3} -\ch{Sr8Ti8O23}, (c) 1 Al-doped, (d) 2 Al-doped \enquote{near}, (e) 2 Al-doped \enquote{near-far}, and (f) 2 Al-doped \enquote{far} oxygen deficient \ch{SrTiO3}. Green, sky blue, red and yellow  balls represent Sr, Ti, O and Al atoms respectively. Projected density of states of (g) pristine (2$\times$2$\times$2) for comparison, (h) \ch{Sr8Ti8O23}, (i) 1 Al-doped, (j) 2 Al-doped \enquote{near}, (k) 2 Al-doped \enquote{near-far} and (l) 2 Al-doped \enquote{far} oxygen deficient \ch{SrTiO3} configurations.}
 \label{figure 2}
\end{figure} 
\subsubsection{3.1.2 Al-doped oxygen deficient \ch{SrTiO3}}
\label{subsubsection 3.1.2}
To the oxygen deficient \ch{SrTiO3}, we add a single Al atom by replacing a Ti atom (Al concentration: 12.5\%). The lowest energy structure is determined by calculating the total energy of 8 different configurations. It is found that energy is lowest for the configuration in which Al atom is close to the oxygen vacancy as shown in Figure 1(c). Next, we replace two Ti atoms by two Al atoms (Al concentration: 25\%). To find the lowest energy configuration, we have considered three different possible scenarios (as done by Zhao \textit{et al.} \cite{Zhao2019}). The naming conventions used by Zhao \textit{et al.} are followed here. In the first case, 2 Al atoms are substituted at the Ti sites near to the oxygen vacancy and this configuration is called the \enquote{near} configuration. The second and third cases correspond to when one of the Al atoms is doped at Ti site near the oxygen vacancy and the other far from it, called the \enquote{near-far} configuration, and two Al atoms are doped far from the oxygen vacancy called the \enquote{far} configuration respectively. In each of these configurations, various combinations are taken into account and total energy for all the combinations are calculated. From the energy calculations, it is found that the \textquote{near} configuration has the lowest total energy when compared to the other two cases [\textquote{near-far} and \textquote{far}] The defect formation energy of 1 Al and 2 Al-doped oxygen deficient \ch{SrTiO3} are discussed in later section and 2 Al-doped is energetically more favorable.\\

The projected density of states (PDOS) of Al-doped oxygen deficient \ch{SrTiO3} are presented in Figure 1(i)-(l). For 1 Al-doped oxygen deficient \ch{SrTiO3} (Figure 1(i)), the in-gap states formed by Ti and O are not fully passivated. For the 2 Al-doped oxygen deficient \ch{SrTiO3} \enquote{near} structure, the vacancy induced states disappear as shown in Figure 1(j) or in other words, Ti$^{3+}$ states formed by oxygen vacancy, are passivated by the substitution of Al$^{3+}$ at Ti sites. The DOS is symmetric with respect to the spin up and spin down part (Figure 1(j)). The VBM is characterized by O-p states and the CBM by Ti-d states for the \enquote{near} configuration. It has an $E_{g}$ of $\sim$ 2.8 eV. The density of states of this \enquote{near} configuration is also calculated using the hybrid functional HSE06 and provided in supporting information (Figure SI4). From the TDOS using both theoretical methods, we see that the VB features are comparable and band gap obtained using HSE06 is $\sim$ 0.2 eV more than the DFT+U method. On applying strain, the band gap of the system is reduced and it lies close to the optimum range for visible light active photocatalysts (details are given in supporting information). For the \enquote{near-far} configuration, the CBM has shifted slightly towards lower energies as there is a localized Ti-O hybrid peak in the CB and the band gap is reduced to 2.50 eV. In the case of \enquote{far} configuration, few in-gap Ti-O states are found in PDOS which further lowers the band gap to $\sim$ 1.76 eV. For \enquote{near-far} and \enquote{far} configurations, the reduced Ti$^{3+}$ states formed by oxygen vacancy, are not fully passivated when compared to \enquote{near} configuration. \\
\subsubsection{3.1.3 (Ir, Al)-codoped oxygen deficient \ch{SrTiO3}}
\label{subsubsection 3.1.3}
\subsubsection*{A. 1:2(Ir, Al)-codoping}
To investigate the effect of Ir codoping on 2 Al-doped oxygen deficient \ch{SrTiO3} (\enquote{near}), we have considered two cationic sites, Sr and Ti, for substitution. Ir is substituted at all possible cationic sites (Ir concentration: 12.5\%) and further calculations are done on the lowest energy structure. The corresponding lowest energy relaxed configuration for Ir substitution at Ti and Sr are presented in Figure 2. Ir being a transition metal has ionic radius closer to that of Ti. At host rich condition, Ir prefers Ti site more than the Sr site (given in supporting information). Since the difference in formation energy (for host rich condition) is not too large and for comparison, we present the DOS of Ir substitution at both cationic sites in supporting information. Also, DFT+U may not be fully sufficient to explain the energetics of the doped systems when compared to advanced hybrid functionals which are computationally demanding \cite{Basera2019}. As can be seen from the relaxed structures in Figure 2, there is greater distortion of atoms when Ir is substituted at Sr site than Ti site, which could be due to the difference in ionic sizes. The large distortion in case of codoping leads to the development of internal field which promotes separation of photo-excited charge carriers and suppression of recombination, thereby enhancing photocatalytic efficiency \cite{Kumar2020}.\\
\begin{figure}
\begin{center}
\subfloat[]{\hspace{-1cm}\includegraphics[scale=0.38]{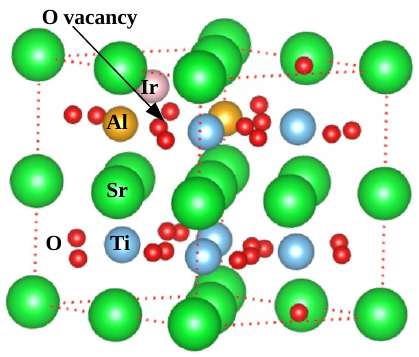}}
\subfloat[]{\hspace{2cm}\includegraphics[scale=0.4]{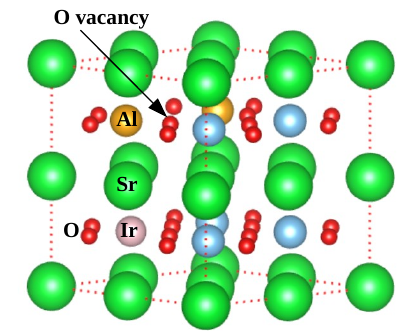}} \\
 \end{center}
 \caption{Relaxed structure of 1:2(Ir,Al)-codoped oxygen deficient \ch{SrTiO3}: (a) Ir substitution at Sr site and (b) Ir substitution at Ti site. Green, sky blue, red, yellow and pink balls represent Sr, Ti, O, Al and Ir atoms respectively.}
 \label{figure 4}
 \end{figure}

The density of states of 1:2(Ir,Al)-codoped oxygen deficient \ch{SrTiO3} are calculated and projected DOS are shown in supporting information (Figure SI6). It is seen that without Hubbard U correction for the Ir-d orbitals, the position of the states are not represented correctly. Here, U$_{d}$(Ir) ranging from 1-6 eV are considered (Figure SI6). Though in-gap hybrid states exist, much larger effective gaps open up with Hubbard correction to Ir-d orbitals, when compared to U$_{d}$(Ir) = 0. The corresponding variation in band gap of 1:2(Ir,Al)-codoped oxygen deficient \ch{SrTiO3} as a function of U$_{d}$(Ir) are represented in supporting information (Figure SI7). The band gap increases with the increase in U and the variation is almost linear. To find the appropriate U value for Ir, we calculate the DOS of 1:2(Ir,Al)-codoped oxygen deficient \ch{SrTiO3} (for Ir substitution at Ti) using HSE06 and compare it with DFT+U results. The U values which gives electronic structure close to that of hybrid functional are selected. For U$_{d}$(Ir) = 3-5 eV, TDOS is given in Figure 3, along with HSE result. We observe that valence band features close to Fermi level obtained with HSE match with that of U$_{d}$(Ir) = 3 whereas the position of in-gap hybrid state complement that of U$_{d}$(Ir) = 4. Therefore, we employ U$_{d}$(Ir) = 3 and 4, for further calculations. The band gap obtained using computationally expensive HSE06 functional is $\sim$ 1.35 eV which lie between the values given by DFT+U for U$_{d}$(Ir) = 3 and 4.\\
\begin{figure}
\centering
\includegraphics[trim=0mm 32mm 0mm 22.5mm,clip,scale=0.8]{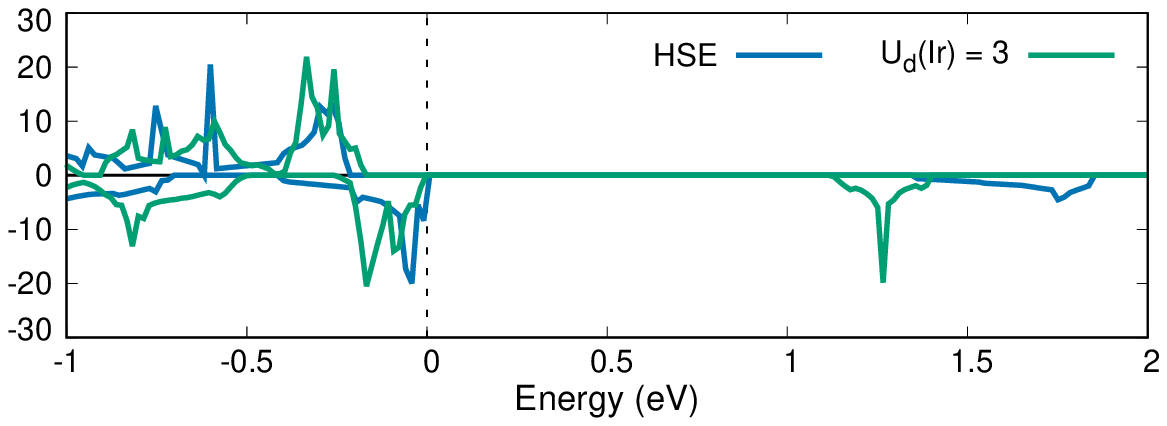} \\
\includegraphics[trim=0mm 32mm 0mm 22.5mm,clip,scale=0.8]{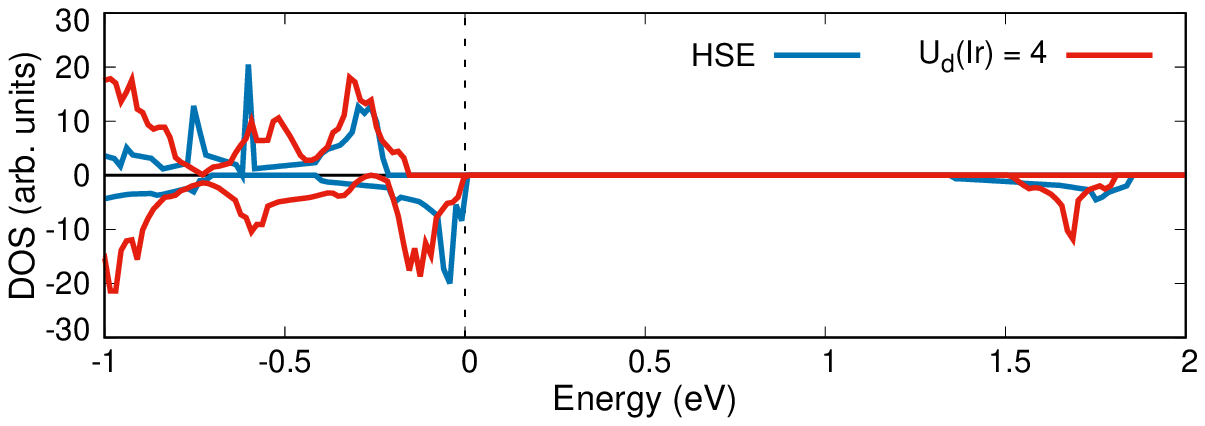} \\
\includegraphics[trim=0mm 18mm 0mm 22.5mm,clip,scale=0.8]{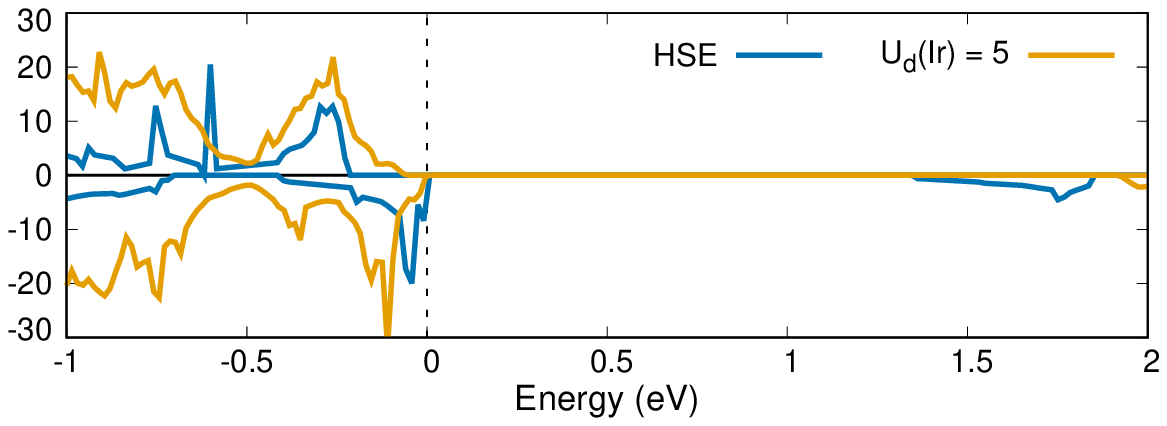} \\
 \caption{Total density of states of 1:2 (Ir,Al)-codoped oxygen deficient \ch{SrTiO3} (Ir substitution at Ti) obtained using DFT+U for U values of 3 - 5 eV in ascending order and HSE06. Fermi level, E${}_{F}$ is represented by the black dotted line at 0 eV.}
 \label{figure 5}
\end{figure}

Figure 4 shows the PDOS of 1:2(Ir,Al)-codoped oxygen deficient \ch{SrTiO3} for U$_{d}$(Ir) = 4. The spin up and spin down part of DOS are unsymmetrical which indicates the presence of unpaired electrons. This is supported by the net magnetic moment of 1 $\mu_B$. From the density of states of Ir d-orbitals (supporting information - Figure SI8) and Bader charge analysis, it can be inferred that Ir is in +4 state in 1:2(Ir,Al)-codoped oxygen deficient \ch{SrTiO3}. The in-gap Ir-O hybrid near the conduction band edge do not form continuous bands with them. Theoretically, the in-gap states can act as charge recombination centers, which reduces the photocatalytic efficiency, albeit visible light absorption. But as in the case of Al-doped \ch{SrTiO3}, if the in-gap states remain unoccupied or partially occupied by electrons due to the shift of Fermi level towards more oxidizing potentials \cite{Zhao2019}, then this system will have improved solar energy conversion properties. Also, it is reported for Ir$^{4+}$:\ch{SrTiO3}, that the presence of nominally unoccupied dopant states in electronic spectrum, only reduced the photo-generated charge collection efficiency, but not make it totally photocatalytically inactive \cite{Kawasaki2014}. It is imperative to verify by X-ray photoelectron spectroscopy (XPS) whether Al-doping reduced the n-type character of Ir-doped \ch{SrTiO3}, which is extensively utilized for photocatalytic \ch{H2} evolution. \\
\begin{figure}
\includegraphics[trim=0mm 32mm 0mm 22.5mm,clip,scale=0.8]{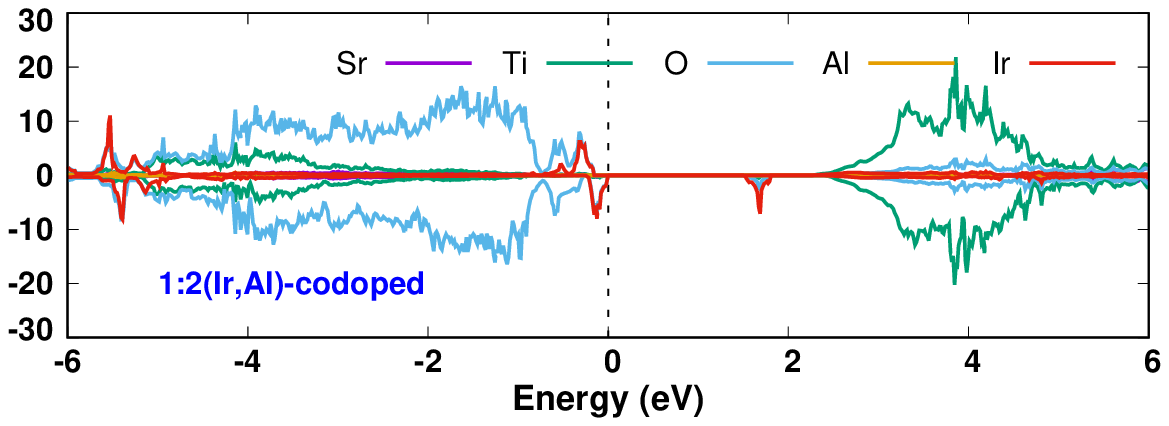} \\
\includegraphics[trim=0mm 32mm 0mm 22.5mm,clip,scale=0.8]{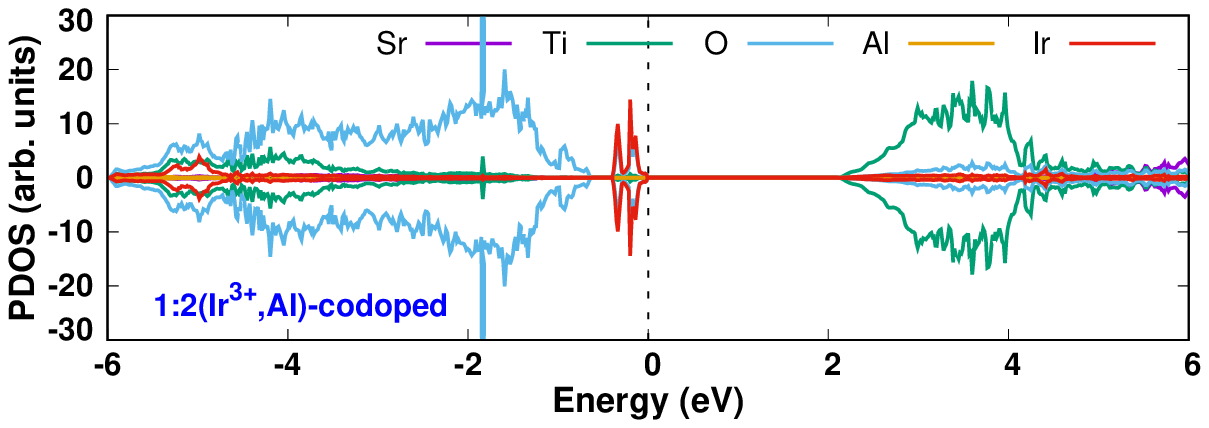} \\
\includegraphics[trim=0mm 18mm 0mm 22.5mm,clip,scale=0.8]{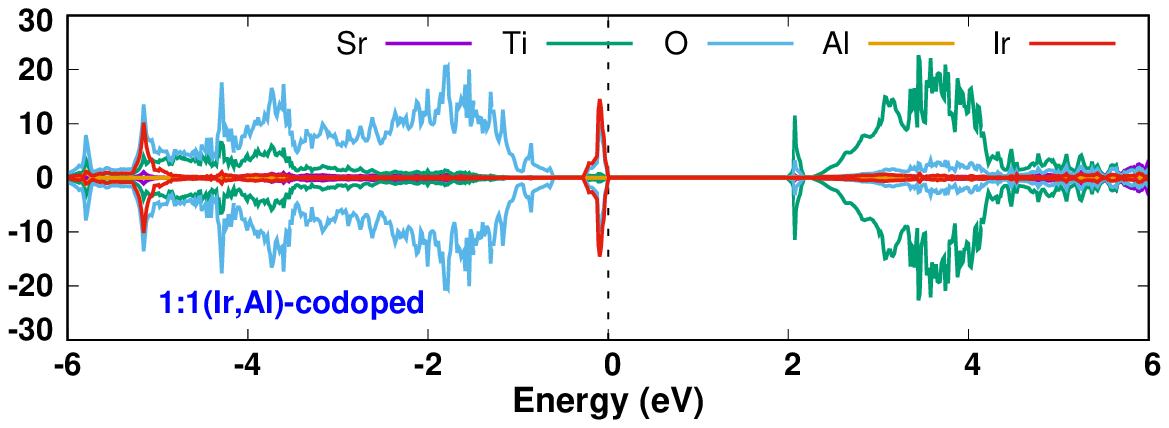} 
 \caption{Projected density of states (PDOS) of 1:2(Ir,Al), 1:2(Ir$^{3+}$,Al) and 1:1(Ir,Al)-codoped oxygen deficient \ch{SrTiO3} for U$_{d}$(Ir) = 4 eV. Fermi level, E${}_{F}$ is represented by the black dotted line at 0 eV.}
 \label{figure 6}
\end{figure}

We have also studied the effect of Ir$^{3+}$ on the electronic structure of 2 Al-doped oxygen deficient \ch{SrTiO3} \enquote{near} configuration. It is modelled by adding an extra electron to 1:2(Ir,Al)-codoped oxygen deficient \ch{SrTiO3}. The PDOS of 1:2(Ir$^{3+}$,Al)-codoped oxygen deficient \ch{SrTiO3} for U$_{d}$(Ir) = 4 is shown in Figure 4. For U$_{d}$(Ir) = 3, the DOS of constituent atoms and Ir d-orbitals are given in supporting information (Figure SI9). The DOS of 1:2(Ir$^{3+}$,Al) looks different from that of 1:2(Ir,Al)-codoped (where Ir is in +4 state), as there are no acceptor (in-gap hybrid) states between Fermi level and conduction band. The VBM is composed mainly of hybridized states of Ir and O while CBM is composed of Ti states. Here, the dopant induced states lie closer to the VB, with a small gap between them. The proximity of dopant induced states to the VB permit electron replenishment and these states may not interfere with CB edge, which determines the driving force for \ch{H2} evolution \cite{Chen2012}. After excitation, if these dopant states are replenished either by absorbing photon or phonon (only for the order of 10 meV), they reduce the probability of trapping charge from CB, hence recombination. In the present case, the gap between O-p and dopant states in the VB is more than 10 meV, therefore the photogenerated holes may not be thermally excited to O-p states, possible only by absorbing photons. The band gap of 1:2(Ir$^{3+}$,Al)-codoped oxygen deficient \ch{SrTiO3} for U$_{d}$(Ir) = 3 and 4 are 1.84 and 2.11 eV respectively, which lies in the optimum range for visible light active photocatalysts. This system has zero net magnetic moment, hence it is charge compensated. From Bader charge analysis, charge on Ir are 1.96 and 1.55 for 1:2(Ir, Al) and 1:2(Ir$^{3+}$,Al)-codoped oxygen deficient \ch{SrTiO3} respectively. This decrease in Bader charge indicates the lower oxidation state of Ir in the latter case. The features in the DOS of 1:2(Ir, Al) and 1:2(Ir$^{3+}$,Al)-codoped oxygen deficient \ch{SrTiO3} obtained in the present work are similar to the HSE results reported by Kawasaki \textit{et al.} \cite{Kawasaki2014} for Ir and Ir$^{3+}$:\ch{SrTiO3} respectively.\\

\subsubsection*{B. 1:1(Ir,Al)-codoped oxygen deficient \ch{SrTiO3}}
We have also investigated the geometry and electronic structure of 1:1(Ir,Al)-codoped oxygen deficient \ch{SrTiO3} (for Ir substitution at Ti site). There are 7 structural configurations and the structure which gives the lowest total energy is considered for further calculations. The net magnetic moment is 0 $\mu_B$ for 1:1(Ir, Al)-codoped and the corresponding PDOS for U$_{d}$(Ir) = 4 is presented in Figure 4. Here, Ir is in the 3+ state (Bader charge is 1.59, similar to the system modelled with an extra electron). In addition to the Ir-O hybrid states at the top of VB, there are Ti-O hybrid states close to the CB. For U$_{d}$(Ir) = 3 and 4, the band gaps are 1.75 and 1.97 eV respectively. Though band gap values are favourable for visible light absorption, there could be some charge recombination depending on the electron occupancy in the in-gap states. This is a charge compensated system as it has zero net magnetic moment. \\

Charge recombination in these systems depend on the electron occupancy in the in-gap states and the position of Fermi level. Therefore, experimental data is needed to complement these DFT results, to further describe the behaviour of (Ir,Al)-codoped \ch{SrTiO3}. 

\subsubsection{3.1.4 Band Edge Alignment}
\label{subsubsection 3.1.4}
Here, we discuss the shift in the band edges of \ch{SrTiO3} with doping and strain. The position of VB and CB edges play an important role in determining the driving force of reduction and oxidation in water splitting process. Specifically, the VBM should be more negative than the oxidation potential of oxygen evolution (\ch{H2O}/\ch{O2}) and CBM, more positive than the reduction potential of hydrogen evolution (H$^{+}$/\ch{H2}). The band edges are calculated based on the relation connecting band gap and electronegativity and is expressed with respect to the absolute vacuum scale as
\begin{equation}
E_C = -\chi + 0.5E_g 
\end{equation}
\begin{equation}
E_V = -\chi - 0.5E_g
\end{equation}
where, $\chi$ is the electronegativity, E$_{g}$ is the band gap of the material, E$_{C}$ and E$_{V}$ are the conduction and valence band edges respectively. For a A$_{a}$B$_{b}$C$_{c}$ compound, the bulk electronegativity can be calculated as the geometric mean of the absolute electronegativities of the constituent elements as
\begin{equation}
\chi = (\chi_A^a \chi_B^b \chi_C^c)^{\nicefrac{1}{a+b+c}}
\end{equation}
The absolute electronegativity of the constituent atoms are taken from the experimental database \cite{Pearson1988}. The valence and conduction band edges of \ch{SrTiO3}, calculated using equations 1 and 2, for DFT+U (HSE) are -6.51 (-6.56) and -3.38 (-3.33) eV respectively. Our results show that VBM of \ch{SrTiO3} is around 0.84 eV more negative than the water oxidation potential and CBM is around 1.06 eV more positive than water reduction potential. The same procedure has been applied for Al-doped and (Ir, Al)-codoped \ch{SrTiO3} to obtain the corresponding band edges. Doping with Al and applying hydrostatic strain, shifts the conduction band edge downwards by a small amount and valence band edge upward for the energetically stable configuration (\enquote{near} - 2 Al atoms doped near the oxygen vacancy). The schematic representation of band edge alignment of Al-monodoped oxygen deficient \ch{SrTiO3} is given in supporting information(Figure SI10). Hence, in theory, the \enquote{near} configuration under strain can act as a photocatalyst for water-splitting. \\
\begin{figure}
\begin{center}
\includegraphics[scale=0.6]{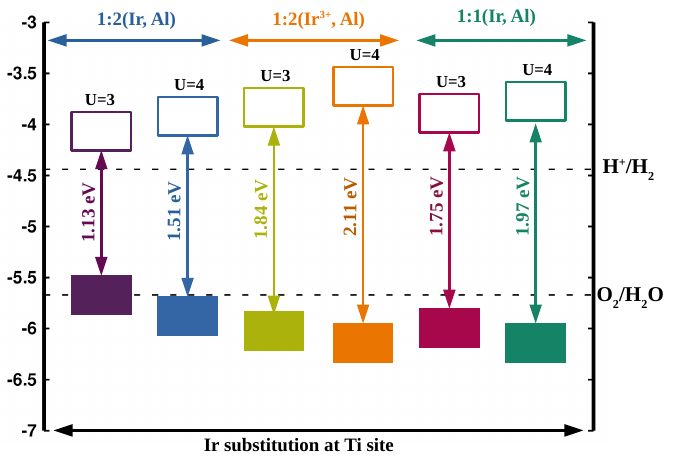}
 \end{center}
 \caption{Schematic representation of band edge alignment of (Ir, Al)-codoped oxygen deficient \ch{SrTiO3} for select U values, with respect to absolute vacuum scale (AVS).  On the electro-chemical scale (NHE), E$_{NHE}$ = -E$_{AVS}$ - 4.50.}
 \label{figure 8}
\end{figure}

The band edge positions obtained using this method gives only a rough estimate as the structural factors are not taken into account \cite{Xu2000}. The schematic representation of band edge alignment of (Ir, Al)-codoped \ch{SrTiO3} for Hubbard potential U$_{d}$(Ir) = 3 and 4 are presented in Figure 5. For 1:2(Ir,Al)-codoped oxygen deficient \ch{SrTiO3}, when U$_d$(Ir) = 4, the valence band edge energy almost coincides with oxidation potential of water which implies it has negligible oxidizing power whereas for 1:2(Ir$^{3+}$,Al)-codoped, the band gap is between 1.8 and 2.2 eV and band edges appear to be at optimum distance from the water redox potentials with greater reducing power which leads to better hydrogen evolution activity. At the same time, for 1:1 (Ir, Al)-codoped system where Ir is in 3+ state, the band edges straddle the water redox potentials with greater driving force for reduction. These results will act as a reference for experimental studies in future.\\

\subsection{3.2 Defect formation energy} 
\label{subsection 3.2} 
The defect formation energy is calculated using the expression
\begin{equation}
E_f = E_{doped} - E_{undoped} + n\mu_{i} - n\mu_{j}
\end{equation}
where $E_{doped}$ is the total energy of the supercell containing dopant/dopants, $E_{undoped}$ is the total energy of the pristine cell of the same size, $n$ is the number of atoms removed or added, and $\mu_{i}$ and $\mu_{j}$ are chemical potentials of the atomic species removed and added respectively. For the calculation of formation energy, the total energies of doped systems and the corresponding chemical potentials of atomic species are calculated using DFT+U corrected potential. Precisely, U$_{d}$ correction is included in the calculation of total energy of Ti/Ir bulk crystal to obtain chemical potential and also U$_{p}$ correction included for calculation of the total energy of oxygen molecule. Such an approach was followed by Park \textit{et al.} to determine the formation energy of oxygen vacancy in TiO$_{2}$ by considering U$_{p}$ correction in the calculation of energy of oxygen molecule \cite{Park2010}. When dopants are added to the oxygen deficient \ch{SrTiO3}, the formation energy equation can be re-written as in equation 5. For example, when two Ti atoms are replaced by two Al atoms in oxygen deficient \ch{SrTiO3}, the formation energy is calculated as
\begin{equation}
E_f = E_{Sr_{8}Ti_{6}Al_{2}O_{23}} - E_{Sr_{8}Ti_{8}O_{24}} + \mu_{O} + 2\mu_{Ti} - 2\mu_{Al}
\end{equation}
Defect formation energy determines the feasibility of synthesis of a doped system. When \ch{SrTiO3} is in equilibrium with reservoirs of Sr, Ti and O, $\mu_{Sr}$ + $\mu_{Ti}$ + 3$\mu_{O}$ = $\mu_{SrTiO_{3}}$. Also, when bulk \ch{SrTiO3} is thermodynamically stable, the heat of formation is calculated as 
\begin{equation}
\Delta = \mu_{SrTiO_{3}(bulk)} - \mu_{Sr(bulk)} - \mu_{Ti(bulk)} - 3\mu_{O(gas)}
\end{equation}
which is negative. The chemical potentials of Sr, Ti and O cannot exceed that of $\mu_{Sr(bulk)}$, $\mu_{Ti(bulk)}$ and $\mu_{O(gas)}$ respectively. Thus $\mu_{Sr}$, $\mu_{Ti}$ and $\mu_{O}$ should satisfy
\begin{align*}
\mu_{Sr(bulk)} + \Delta \leq \mu_{Sr} \leq \mu_{Sr(bulk)}\\
\mu_{Ti(bulk)} + \Delta \leq \mu_{Ti} \leq \mu_{Ti(bulk)} \\
3\mu_{O(gas)} + \Delta \leq 3\mu_{O} \leq 3\mu_{O(gas)}
\end{align*}   

Figure 6 presents the variation of defect formation energy of Al-monodoped, (Ir, Al)-codoped \ch{SrTiO3} for Ir substitution at Ti site as a function of chemical potential ($\mu_{Ti}$ - $\mu_{Ti(bulk)}$). It is seen that the defect formation energy increases when going from Ti-poor to Ti-rich conditions. This is expected for substitutional doping as vacancies are necessary and they are difficult to form under host-rich condition. From Figure 6, it is seen that 2 Al atom substitution at Ti is energetically more favourable than single Al atom doping. For Ir substitution at Sr site in 1:2(Ir,Al)-codoping, the change in defect formation energy as a function of chemical potential is given in supporting information. At host rich conditions, formation energy for Ir substitution at Ti is lower than that of Ir substitution at Sr in 1:2(Ir:Al) codoping (though the difference is small). Formation energy calculation also indicates that 1:2(Ir,Al)-codoping is energetically more favourable than 1:1(Ir,Al)-codoping. When shifting away from Ti-rich condition, the cost of 1:2(Ir,Al)-codoping becomes less than Al-monodoping (2 Al-doped) in oxygen deficient \ch{SrTiO3}.
\begin{figure}
\begin{center}
\includegraphics[scale=0.7]{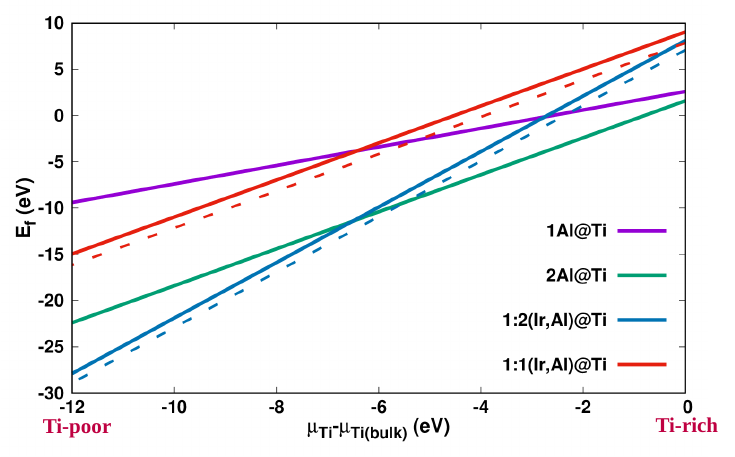} 
 \end{center}
 \caption{Variation of defect formation energy as a function of chemical potential of Ti($\mu_{Ti}$ - $\mu_{Ti(bulk)}$) for Al-monodoped and (Ir,Al)-codoped oxygen deficient \ch{SrTiO3}. The continuous colored lines correspond to U$_{d}$(Ir) = 0 and dashed colored lines represent U$_{d}$(Ir) = 4 in (Ir,Al)-codoped systems.}
 \label{figure 9}
\end{figure}

\subsection{3.3 Optical properties}
\label{subsection 3.3}
To determine whether the absorption peaks of the engineered compounds fall in the visible region, we have calculated the optical properties. The frequency dependent dielectric function has been calculated in the independent particle approximation and complex dielectric function is represented as $\varepsilon(\omega)$ = $\varepsilon_{1}(\omega)$ + $i\varepsilon_{2}(\omega)$, where $\varepsilon_{1}(\omega)$ and $\varepsilon_{2}(\omega)$ are the real and imaginary parts of dielectric function respectively. The real part is calculated from the imaginary part using Kramers-Kronig relation. The determination of optical properties using hybrid functional is cumbersome, hence we use DFT+U to obtain a profile of the solar light absorption in codoped systems considered here. The absorption curves of Al-doped and strained oxygen deficient \ch{SrTiO3} are provided in supporting information (Figure SI12(a)). The strained systems have better absorption in the 350-450 nm wavelength region when compared to pristine \ch{SrTiO3}. The absorption coefficient as a function of wavelength of codoped systems for selected U values are presented in Figure 7. 
For U$_{d}$(Ir) = 3 and 4, 1:2(Ir, Al)-codoped system has better visible light absorption as it extends over a wide range of wavelength. But, its electronic structure has induced dopant-O hybrid level in the forbidden region, which might reduce the amount of excited charge carriers. On reducing Ir$^{4+}$ in 1:2(Ir, Al)-codoped system to Ir$^{3+}$ by adding an electron, the optical absorption spectrum obtained is different. For the same, the absorption spectrum for U$_{d}$(Ir) = 3 and 4 are presented in Figure 7. There is absorption in the 350-550 nm wavelength region for both U values and negligible beyond 550 nm. \\
\begin{figure}
\begin{center}
\includegraphics[scale=0.8]{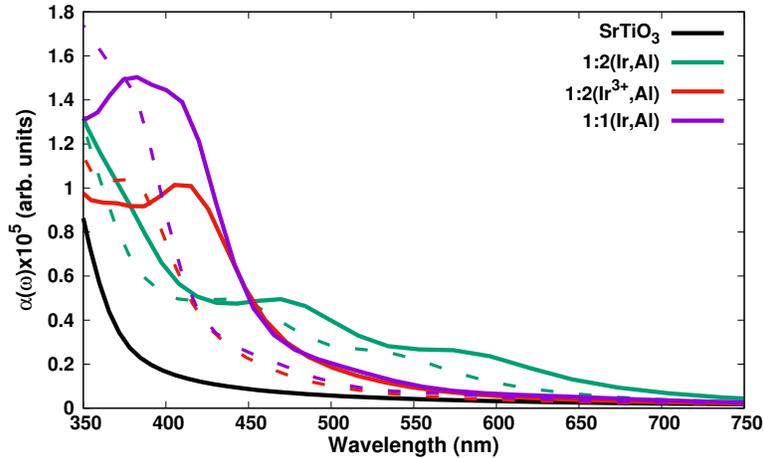} 
 \end{center}
 \caption{Absorption coefficient as a function of wavelength for (Ir,Al)-codoped oxygen deficient \ch{SrTiO3}. Here, the continuous colored lines correspond to U$_{d}$(Ir) = 3 and dashed colored lines represent U$_{d}$(Ir) = 4.}
 \label{figure 10}
\end{figure}

Figure 7 also shows the absorption curve of 1:1(Ir, Al)-codoped oxygen deficient \ch{SrTiO3} as a function of wavelength for U$_{d}$(Ir) = 3 and 4. This is similar to the case of 1:2(Ir$^{3+}$,Al)-codoped oxygen deficient \ch{SrTiO3}, but with higher absorption in the 350 - 440 nm wavelength region. For higher values of $\lambda$, curves corresponding to both 1:2(Ir$^{3+}$,Al) and 1:1(Ir, Al)-codoped coincide, which again affirms that they have same oxidation state because of the similarity in optical signature. Based on absorption spectrums, 1:2(Ir, Al)-codoped has a wider optical response in visible region whereas 1:2(Ir$^{3+}$,Al)-codoped and 1:1(Ir, Al)-codoped oxygen deficient \ch{SrTiO3} have better response in the low wavelength region (of visible light spectrum). 

\section{4. Conclusions}
\label{section 4}
In the present study, using DFT+U method, we have investigated the effect of Al-monodoping and (Ir,Al)-codoping on the electronic structure and optical properties of oxygen deficient \ch{SrTiO3}. The outcome of codoping Al and Ir in different ratios on \ch{SrTiO3} depend on competing factors like position of dopant induced states with respect to Fermi level and their occupancy (determine probability of charge recombination), band edge alignment with respect to water redox levels (give an idea of driving force for oxidation and reduction) and visible light absorption ability (based on the decrement in band gap). The smaller the band gap values, higher the light absorption and too small gaps lead to band edges that do not straddle reduction and oxidation levels of water. The position of Ir-O hybrid states in the forbidden region of 1:2(Ir,Al)-codoped oxygen deficient \ch{SrTiO3}, can cause a certain amount of photo-excited charge carriers to recombine, which reduces overall efficiency. This depends on the electron occupancy in these states. 1:2(Ir$^{3+}$,Al)-codoped system has no acceptor states between Fermi level and CB, with higher absorption in the 350 - 460 nm region than 1:2(Ir,Al)-codoped. For 1:1(Ir,Al)-codoped, absorption curves coincide with 1:2(Ir$^{3+}$,Al)-codoped oxygen deficient \ch{SrTiO3} beyond 410 nm. The position of Fermi level and the occupancy of dopant induced states (thereby probability of charge recombination) in (Ir,Al)-codoped \ch{SrTiO3} needs to be deduced from experiments in combination with these ab-initio results.  Therefore, the exact behaviour of these materials can be further understood with complimentary experiments (effect of temperature and processing conditions) and/or DFT calculations with hybrid functionals, that would provide rational insights towards designing efficient photocatalysts.

\begin{suppinfo}
\label{supporting_information}
The band structure of pristine \ch{SrTiO3} obtained using GGA and DFT+U, DOS of pristine and 2 Al-doped oxygen deficient \ch{SrTiO3} using hybrid functional, details of effect of strain on Al-doped oxygen deficient \ch{SrTiO3}, projected DOS of codoped and 3$\times$3$\times$3 cell, formation energy plot and absorption spectra of strained 2 Al-doped \ch{SrTiO3} and 1:2(Ir,Al)-codoped system (for Ir substitution at Sr) are provided in Supporting Information.
\end{suppinfo}

\bibliography{STO}

\end{document}